\documentclass[twocolumn, trackchanges]{aastex701}
\usepackage{amsmath}
\usepackage{newtxtext,newtxmath}
\usepackage{xcolor}
\usepackage{adjustbox}
\usepackage{graphicx}
\usepackage{amsmath}
\usepackage{footnote}
\usepackage{amssymb}

\def \rxte{{\it RXTE}}
\def \inte {{\it INTEGRAL}}

\def \xmm {{\it XMM}-Newton}
\def \src{{4U~1702--429}}

\def \nustar{{\it NuSTAR}}

\def \nicer{{\it NICER}}

\def \astrosat{\textit{AstroSat}}

\def \relxill{{\tt relxill}}

\newcommand{\erg}{erg cm$^{-2}$ s$^{-1}$} 
\newcommand{\lum}{erg s$^{-1}$}

\shorttitle{Photospheric radius expansion burst and disk reflection from \src{}}
\shortauthors{Mandal et al.}

\begin{document}

\title{Photospheric radius expansion thermonuclear burst and X-ray reflection from the neutron star X-ray binary \src{}}

\correspondingauthor{Manoj Mandal}
\email{manojmandal213@gmail.com}

\author[orcid=0000-0002-1894-9084]{Manoj Mandal}
\affiliation{Astronomy and Astrophysics Division, Physical Research Laboratory, Navrangpura, Ahmedabad - 380009, Gujarat, India}
\email{manojmandal213@gmail.com}

\author[orcid=0000-0003-2865-4666]{Sachindra Naik} 
\affiliation{Astronomy and Astrophysics Division, Physical Research Laboratory, Navrangpura, Ahmedabad - 380009, Gujarat, India}
\email{snaik@prl.res.in}

\author[orcid=0000-0002-6789-2723]{Gaurava Kumar Jaisawal} 
\affiliation{DTU Space, Technical University of Denmark, Elektrovej 327-328, DK-2800 Lyngby, Denmark}
\email{gaurava@space.dtu.dk}

\begin{abstract}
We perform a comprehensive study of thermonuclear bursts from the neutron star low-mass X-ray binary \src~ detected with \nicer~ and \xmm. The thermonuclear burst detected with \nicer~ shows clear evidence of a photospheric radius expansion (PRE) event and a distinct feature in the burst profile. The burst profiles demonstrate significant energy dependence, with the hardness ratio varying notably during the PRE phase. The radius of the neutron star photosphere expanded to a maximum of $23.1_{-3.2}^{+3.8}$~km while its temperature reached a minimum of $\sim1.4$ keV. The time-resolved burst spectra can be modeled using variable persistent emission method, indicating that the soft excess may arise from enhanced mass accretion onto the neutron star, potentially due to the Poynting-Robertson drag. Alternatively, the disk reflection model can be used to explain the soft excess emission during a burst. The time-resolved spectral study is performed for three thermonuclear bursts detected with \xmm. The \xmm{} time-resolved burst spectra can be modeled using an absorbed blackbody model, without any signatures of the PRE. We conduct a detailed spectral analysis of the 2025 \nustar{} observation of \src, revealing a broad iron line at 6.4 keV and a Compton hump around 20 keV, indicating X-ray reflection features. The disk reflection model {\tt relxill} provides an inner disk radius of $\sim$12~\(R_g\) and an inclination angle of $\sim39^{\circ}$. The magnetic field strength at the pole of the neutron star is estimated to be \(5.1~\times~10^{8}\)~G, assuming that the accretion disk is truncated at magnetosphere boundary.
\end{abstract}


\section{Introduction}
\label{intro}
The neutron star low-mass X-ray binary (LMXB) system contains a neutron star (NS) and a low mass ($\le 1 M_\odot$) companion star orbiting around the common center of mass. In an LMXB system, the NS is comparatively weakly magnetized with a field strength in the range of $B\sim10^7-10^9$ G \citep{Ca09, Mu15}. In the case of LMXBs, matter gets accreted onto the NS from the companion star through the Roche lobe overflow \citep{Fr02}. NS-LMXBs can be categorized into two groups, known as `atoll' and `Z' sources, based on the shape of the track in the hardness-intensity diagram (HID) and the color-color diagram (CCD) \citep{Ha89, Va04}. The $Z$ sources are comparatively brighter ($L>0.5L_{\textrm{Edd}}$, $L_{\textrm{Edd}}$ stands for Eddington luminosity) than the atoll sources ($L<0.5L_{\textrm{Edd}}$) \citep{Do07}. The $Z$ and atoll sources also exhibit different timing and spectral properties \citep{Va06}.

The LMXBs show thermonuclear bursts, which occur because of the unstable burning of the accreted matter, hydrogen/helium, on the surface of the NS. The lighter elements transform to heavier elements through nuclear chain reaction during the burst \citep{Le93, St03, Sc06}. The burst exhibits a rapid increase in X-ray luminosity over a few seconds and is followed by an exponential decay with a decay time of tens of seconds \citep{Le93}. Burst emission is typically described using a blackbody model by assuming that the NS emits like a perfect blackbody \citep{Va78, Ku03}. In some cases, the burst peak luminosity can reach the Eddington limit, which may affect the environment of the disk \citep{De18}, causing deviation from a pure blackbody emission for different LMXBs, e.g., Aql~X-1 \citep{Ke18a, Gu22, Ma25}, 4U~1820-30 \citep{Ke18b, Ja24}, 4U~1608-52 \citep{Ja19}, etc. A soft and hard excess can be observed in the spectrum that can be characterized using the variable persistent emission method/$f_a$ method \citep{Worpel2013}. Alternatively, the disk reflection model can be used to explain excess emission, implying that a fraction of burst photons may interact with the accretion disk and be reprocessed and reflected from it \citep{Zh22, Lu23}. When the peak luminosity reaches the Eddington limit, the radiation pressure is high enough to exceed the gravitational pull, and the photosphere expands over the surface of the NS, resulting in a photospheric radius expansion (PRE) event. At a constant luminosity level near the burst peak, the radius of the apparent emitting region increases significantly, and the temperature of the photosphere drops to its minimum \citep{Ku03}. This effect is also reflected as a double-peak profile in the light curve above 3 keV. In some cases of intense PRE bursts, the photospheric radius increases by 10-100 km \citep{Ga08}. In case of a super-expansion burst, the photosphere can reach up to 1000 km \citep{in10}. The compactness of the NS can be constrained by tracing the evolution of the spectral parameters or by detecting the burst-wind features during PRE bursts \citep{va87, Le93, Ku03, Ja25}.

The NS-LMXB \src{} (Ara~X-1) was discovered in 1976 with the 8th Orbiting Solar Observatory (OSO-8) by \citet{Sw76}. It was later classified as an atoll source based on observations from {\it EXOSAT} \citep{Oo91}. Thermonuclear X-ray bursts from \src{} have been observed with \rxte{} \citep{Ma99}, \xmm{} \citep{Ia16} and \astrosat{} \citep{Va24} observatories. The source is also known to show burst oscillations at 330 Hz with \rxte{} \citep{Ma99}, suggesting the spin frequency of the NS \citep{Ch03}. In addition, \citet{Ma99} reported kilohertz quasi-periodic oscillation (QPO) from \src{} using \rxte{} observation.

Using PRE bursts observed with \rxte{}, the distance to \src{} was estimated to be about 4.2 kpc assuming a hydrogen-rich companion, or 5.5 kpc for helium \citep{Ga08}. Previous \xmm{} and \inte{} observations primarily focused on reflection features from the persistent emission but did not analyze bursts detected during \xmm{} observation \citep{Ia16}. Reflection features in NS-LMXB spectra arise when hard Comptonized photons scatter off the accretion disk, producing a Compton hump around 20–40 keV and fluorescent emission lines, notably a broad iron K$\alpha$ line at 6.4–6.97 keV \citep{Fa89, Fa00, Mi07, Ca08}. The iron line’s broadening results from Doppler effects, scattering in the inner disk, and relativistic effects near the neutron star \citep{Fa89, Fa00, Mi07}. Since these features originate close to the neutron star, reflection modeling is a powerful tool for probing disk properties and, assuming disk truncation near the magnetosphere, can also constrain the magnetic field of the neutron star \citep{Ca09, Lu19}. Recent studies with \nicer{} and \astrosat{} have also examined timing and spectral properties of the persistent emission, detecting an 800 Hz QPO \citep{Ch24}. Time-resolved spectroscopy of three bursts with \astrosat{}/LAXPC revealed photospheric radius expansion ($\sim$25 km) in one burst. However, no excess emission was detected during PRE bursts, likely due to the limited soft X-ray sensitivity of \astrosat{}/LAXPC \citep{Va24}. This underscores the importance of further soft X-ray studies with instruments like \nicer{} to better understand burst emission and burst-disk interactions.

This work presents a comprehensive study of thermonuclear X-ray bursts from \src{} using soft X-ray observations from \nicer{} and \xmm{}. A PRE burst with a distinctive profile is detected for the first time with \nicer{}, and a detailed time-resolved spectral analysis is conducted to examine the evolution of emission components throughout the burst. The relativistic disk reflection model is employed to investigate burst-disk interactions and variations in disk properties. The bursting regime and underlying burning processes are further characterized by estimating the mass accretion rate, ignition column depth, and burst fluence. To complement this analysis, three additional bursts observed with \xmm{} EPIC-PN were examined for spectral behavior. We also studied broadband persistent emissions using \nustar{} observation from March 2025, with reflection modeling via the {\tt relxill} framework used to constrain accretion disk structure and estimate the magnetic field strength of the neutron star.
The paper is organized as follows: Section~\ref{obs} describes observations and data reduction; Section~\ref{res} presents spectral and timing results; Section~\ref{dis} discusses implications; and Section~\ref{con} summarizes our conclusions.

  \begin{figure}
\centering{
\includegraphics[width=\columnwidth]{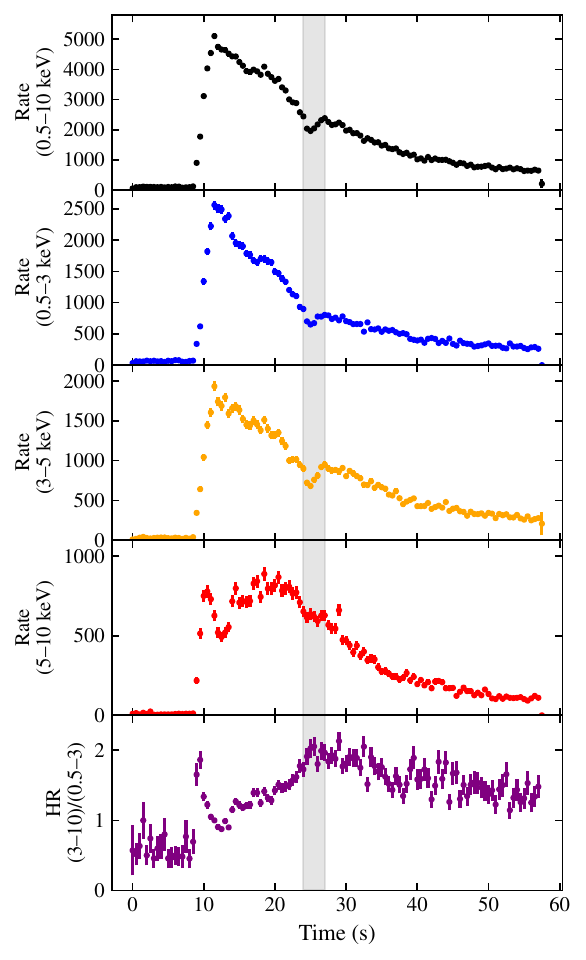}
\caption{{The light curves of \src{} are shown for 4 different energy ranges using \nicer{} observation. The bottom panel represents the evolution of the hardness ratio (HR) during the thermonuclear burst. The shaded region indicates a weak secondary peak in the burst light curve.} 
}
\label{fig:HR}}
\end{figure}

\begin{table*}
\centering
\caption{Summary of observations of \src{} using \nicer{}, \xmm{}, and \nustar{}.}
\begin{tabular}{lccc} 
\hline
Observatories & Date of Observations &  Observation ID & Exposure (ks)  \\
\hline
\xmm{} &  2010-03-09 (MJD 55264) &   0604030101 &  38     \\
\nicer{} &  2019-08-09 (MJD 58704) &  2587030102  &  18      \\
\nustar{} &  2025-03-21 (MJD 60755) &   91101305002 &  23     \\

\hline
\label{tab:log_table_burst}
	\end{tabular}
\end{table*}

\section{Data analysis and methodology}
\label{obs}
In this work, we used publicly available \nicer{}~ and \xmm{} data for which a total of 4 thermonuclear X-ray bursts are detected. In addition, we also used data from our recent \nustar{} Target of Opportunity (ToO) observation in March 2025 during an outburst to investigate the X-ray reflection feature in the source spectrum. The observation details are summarized in Table~\ref{tab:log_table_burst}. No X-ray burst was detected during the \nustar{} observation.

\subsection{\nicer{} observation}
The Neutron Star Interior Composition Explorer (\nicer{}) onboard at the International Space Station (ISS) is a non-imaging telescope, operating in the soft X-ray energy range of 0.2-12 keV \citep{Ge16}. \nicer{} monitored \src{} between 2017-2019. In this work, we focus on the observation in August 2019 during which a thermonuclear burst was detected. The details of the \nicer{} observation are mentioned in Table~\ref{tab:log_table_burst}. Using the {\tt NICERDAS} tool in {\tt HEASOFT}, we process the raw data. Clean event files are created using the calibration tool {\tt nicerl2}. To analyze the \nicer{} data, we use {\tt CALDB} version xti20240206. The {\tt barycorr} tool is utilized to apply barycentric correction to the data. The light curves and spectra are generated using the {\tt XSELECT} package. The tools {\tt nicerrmf} and {\tt nicerarf} are used to build the response and auxiliary response files for further spectral analysis. The {\tt nibackgen3C50} tool is used to create the background spectrum \citep{Re22}.

\subsection{\xmm{} observation}
The European Photon Imaging Camera (EPIC, 0.1-15 keV) is mounted at the focus of each of the three 1500 cm$^2$ X-ray telescopes of the \xmm{} observatory \citep{Ja01}. One of the EPIC imaging spectrometers utilizes advanced PN charge-coupled devices (CCDs) \citep{St01}, while the other two use MOS CCDs \citep{Tu01}. The reflection grating spectrometers (RGS), operating at 0.35-2.5 keV, are placed behind two telescopes \citep{He01}. \src{} was observed with the \xmm{} observatory on March 9, 2010, with an exposure of $\sim$38 ks, and we used the EPIC-PN timing mode data in this work. The \xmm{} Science Analysis System ({\tt SAS}) version 20.0.0 is used to reduce the raw XMM-Newton data. The EPIC-PN  event files are generated using {\tt epproc}. The events are filtered using the {\tt evselect} routine with the criteria PATTERN $\le$4 and FLAG = 0. The source events are generated from a 24-pixel-wide strip centered at the source position RAWX = 36 (i.e., RAWX in [24:48]), while the background region is chosen away from the source in RAWX columns [2:12]. The {\tt evselect} routine is used to generate the final product for the source and background light curves and spectra. Subsequently, background subtraction and corrections are performed using {\tt epiclccorr}. The tasks {\tt rmfgen} and {\tt arfgen} are used to generate spectral response files for each spectrum.
 
\subsection{\nustar{} observation}
The Nuclear Spectroscopic Telescope Array (\nustar{}) is an imaging X-ray telescope made up of two identical co-aligned detectors (FPMA \& FPMB) operating in the 3–79 keV energy range \citep{Ha13}. We proposed a \nustar{} Director’s Discretionary Time (DDT) observation of \src, which was conducted on 21 March 2025 for an exposure of $\sim$23 ks. The observation details are summarized in Table~\ref{tab:log_table_burst}. The standard \nustar{} data analysis tool {\tt NUSTARDAS}, which comes with {\tt HEASOFT}, and the  {\tt CALDB} version 20240325 are utilized for data reduction. The event files are filtered using the {\tt nupipeline} task. To extract the source events, a circular region with a 100 arcsecond radius, centered on the source position, is selected. To generate background events, an identical circular region is defined away from the source region. The light curve, energy spectrum, auxiliary response file, and response matrix file for both detectors are subsequently produced using the {\tt nuproducts} task. The background correction of the light curves is carried out with the {\tt lcmath} task.  The spectrum from each detector is grouped for a minimum of 25 counts bin$^{-1}$ using {\tt grppha}. Further, we simultaneously model the energy spectra from both detectors in {\tt XSPEC} \citep{Ar96} to minimize the cross-calibration uncertainties.  

\section{Results}
\label{res}
In this work, we conduct a detailed study of four thermonuclear bursts detected with \nicer{} and \xmm{} from \src{}. The X-ray burst recorded by \nicer{} exhibits a secondary peak-like feature in the burst profile. We also present the results from the spectral analysis of the persistent emission based on a recent \nustar{} observation conducted in 2025. During this \nustar{} observation, no thermonuclear burst was detected. Additionally, we observe an X-ray reflection feature in the persistent emission, which is further analyzed using the relativistic disk reflection model.

\begin{table}
\centering
 \caption{Best-fitting pre-burst spectral parameters of \src{} with the {\tt XSPEC} model $\texttt{TBabs}\times(\texttt{po}\,+  \texttt{bbodyrad})$ using \nicer{} and \xmm{}.}
\begin{tabular}{llccc}
\hline
Parameters & \nicer{} & \multicolumn{3}{c}{\xmm{}} \\
  &  & TNB1 & TNB2 & TNB3  \\
\hline
N$_{\textrm{H}}$ & 2.4$\pm0.1$ & $2.3\pm0.1$ & $2.2\pm0.1$ & $2.1\pm0.1$  \\
\hline
 $\Gamma$ & 1.9$\pm0.1$ & $2.0\pm0.1$ & $1.8\pm0.1$ & $1.8\pm0.1$  \\
 Norm. & 0.15$\pm0.01$ & $0.5\pm0.03$ & $0.4\pm0.02$ & $0.4\pm0.02$  \\
  \hline
kT$_{\textrm{bb}}$(keV) & 1.2$\pm$0.1 & $1.5\pm0.1$ & $1.5\pm0.1$ & $1.4\pm0.1$ \\
Norm. & 11.0$\pm$2.0 & $11.5\pm1.0$ & $10.0\pm1.2$ & $13.0\pm1.0$ \\
\hline
Flux$_{\textrm{Total}}$  & 2.40$\pm 0.05$ & $6.6\pm0.04$ & $7.0\pm0.04$ & $6.5\pm0.03$  \\
Luminosity  & 0.9$\pm$ 0.1 & $2.5\pm0.1$ & $2.7\pm0.1$ & $2.5\pm0.1$  \\
\hline
$\chi^{2}$/dof & 747/666 & 2007/1688 & 1887/1702 & 1934/1671  \\
\hline
\label{tab:tab_preburst}
\end{tabular}\\
{\bf $f$} indicates the frozen parameters. 
{\bf }: The reported unabsorbed fluxes are in the energy range of 0.1-100 keV and in unit of 10$^{-9}$ ergs cm$^{-2}$ s$^{-1}$, the X-ray luminosity is in the units of 10$^{37}$ erg s$^{-1}$, $N_H$ is in the unit of $\rm 10^{22}~ cm^{-2}$.
\end{table}

\subsection{Burst light curves and hardness ratio}
We report a detailed study of a burst detected with \nicer{} during the August 2019 observation. The burst light curve shows a peak count rate of $\sim$5100 counts s$^{-1}$ in the 0.5-10 keV energy band. Figure~\ref{fig:HR} displays the light curves at various energy ranges during the burst. The figure shows a strong energy dependence of the burst profile. The peak count rate in the 0.5-3 keV energy band is $\sim$2500 counts s$^{-1}$, which decreased to $\sim$800 counts s$^{-1}$ in the 5-10 keV energy band. The burst profile also evolved significantly with energy. A hint of a weak secondary peak, marked with a shaded vertical strip in Figure~\ref{fig:HR}, is also visible in the burst light curve. The secondary peak is less prominent in low-energy bands (below 3 keV). The hardness ratio (HR), the ratio between the light curves in the 3-10 keV and 0.5-3.0 keV ranges, evolved significantly during the burst. The bottom panel of Figure~\ref{fig:HR} shows the evolution of HR as the burst progressed. During the peak of the burst, the HR attains its minima, indicating that the soft photons dominate over the hard photons. 

\begin{figure*}
\includegraphics[width=0.65\columnwidth]{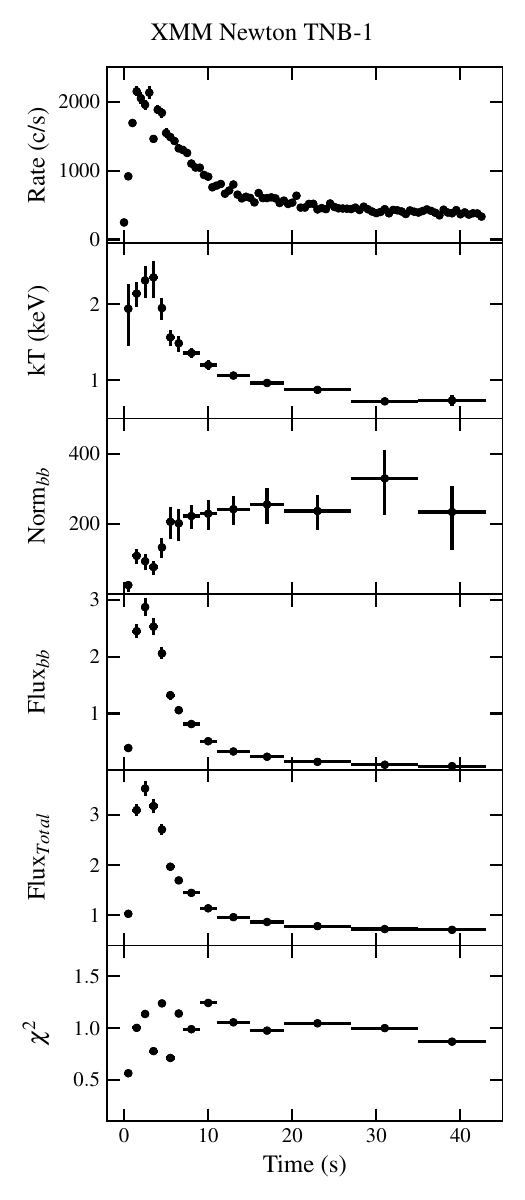} 
\includegraphics[width=0.65\columnwidth]{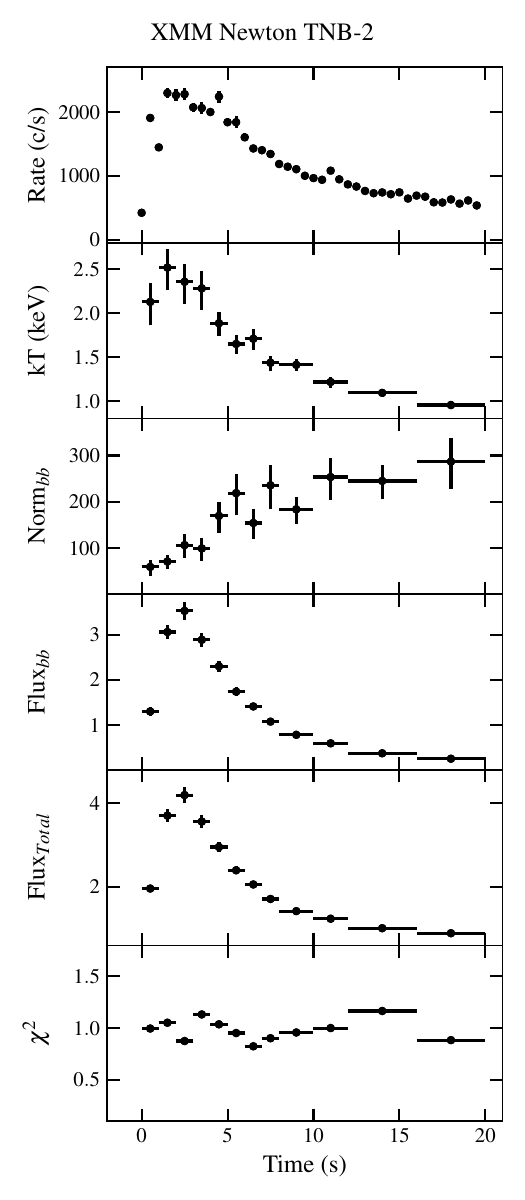}
\includegraphics[width=0.65\columnwidth]{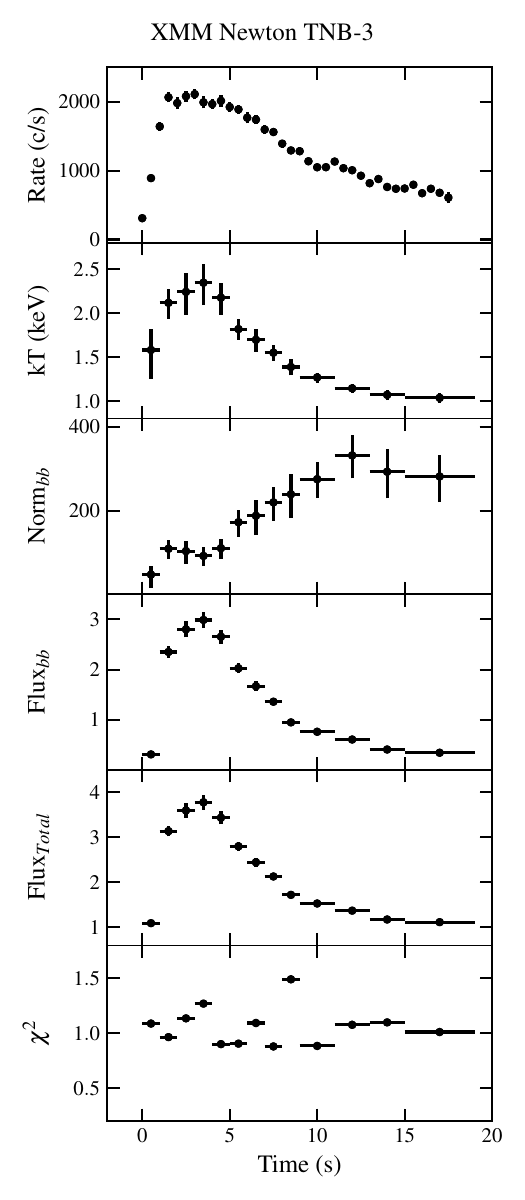}   
\caption{Evolution of spectral parameters of \src\ obtained with the spectral modeling of data with the absorbed blackbody model during three thermonuclear bursts observed with \xmm{} EPIC-PN. The reported flux values are in the unit of 10$^{-8}$ erg cm$^{-2}$ s$^{-1}$. The top panel shows the 0.5-s binned EPIC-PN light curve during the burst.}
    \label{fig:fig_time_Resolved_XMM}
\end{figure*}

The burst profile and the light curve in 0.5-10 keV are also generated for the \xmm{} EPIC-PN data. The 0.5~s binned EPIC-PN light curves for three bursts are shown in the top panels of Figure~\ref{fig:fig_time_Resolved_XMM}. The peak count rates (0.5-10 keV) for EPIC-PN TNB-1, TNB-2, and TNB-3 are $\sim$2200 counts s$^{-1}$, $\sim$2300 counts s$^{-1}$, and $\sim$2000 counts s$^{-1}$, respectively.

\begin{figure*}
     \includegraphics[width=0.92\columnwidth]{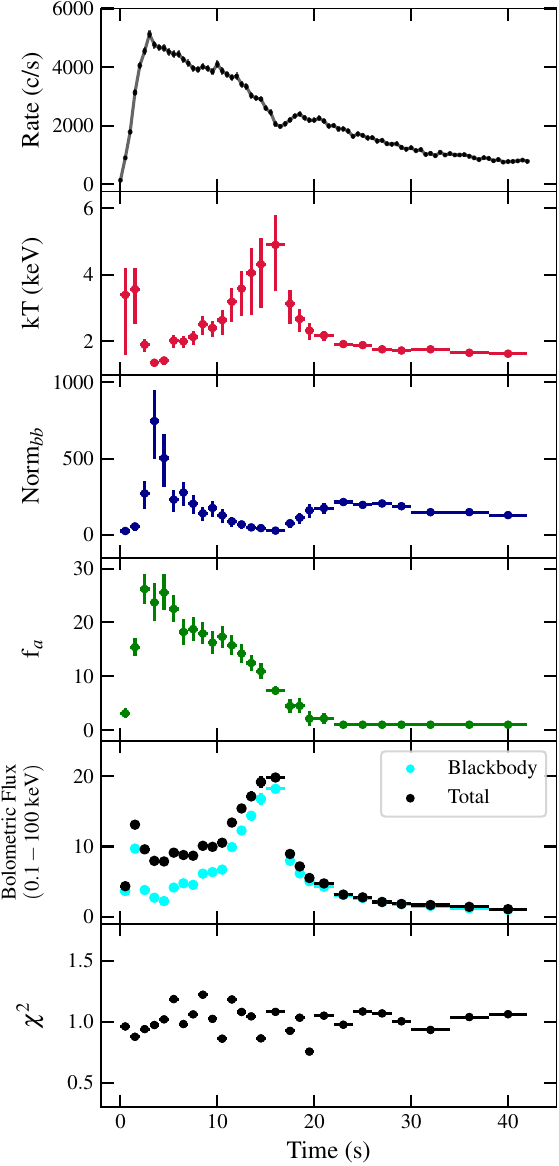}
  \hspace{0.02\textwidth}
  \includegraphics[width=0.92\columnwidth]{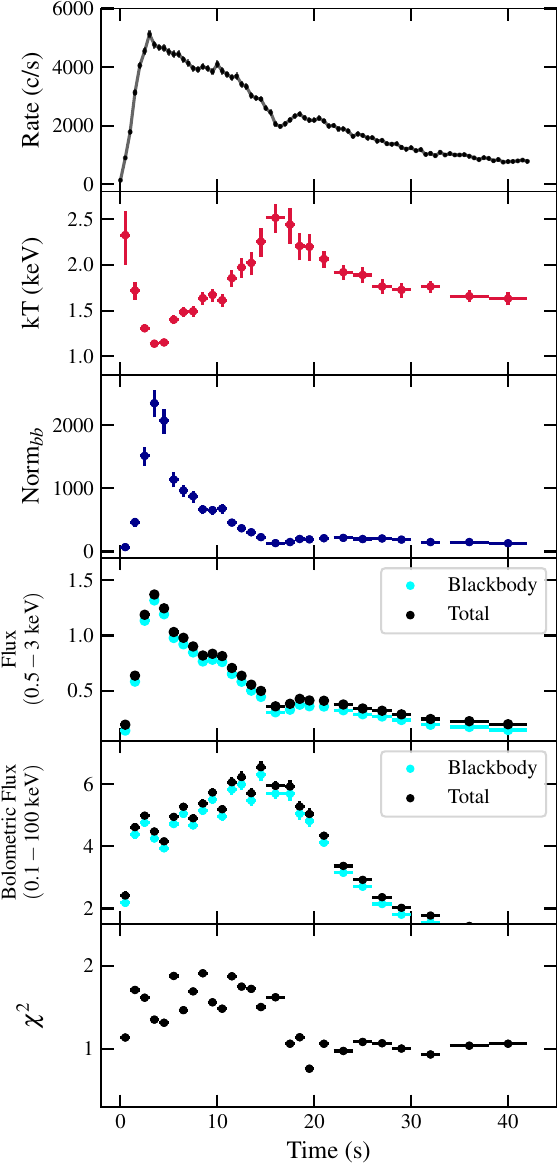}
\caption{Evolution of different spectral parameters of \src\ obtained from the spectral modeling of the data with the variable persistent emission method ($f_a$ method; left panel) and without the $f_a$ method (right panel) during the thermonuclear burst detected with \nicer{}. The top panel shows the 0.5-s binned \nicer{} burst light curve in the 0.5-10 keV energy band. The reported flux values are in the unit of 10$^{-8}$ erg cm$^{-2}$ s$^{-1}$. The expansion of the NS photosphere is observed during the burst.}
    \label{fig:fig_time_Resolved_nicer}
\end{figure*}
\begin{figure}
 \includegraphics[width=0.92\columnwidth]{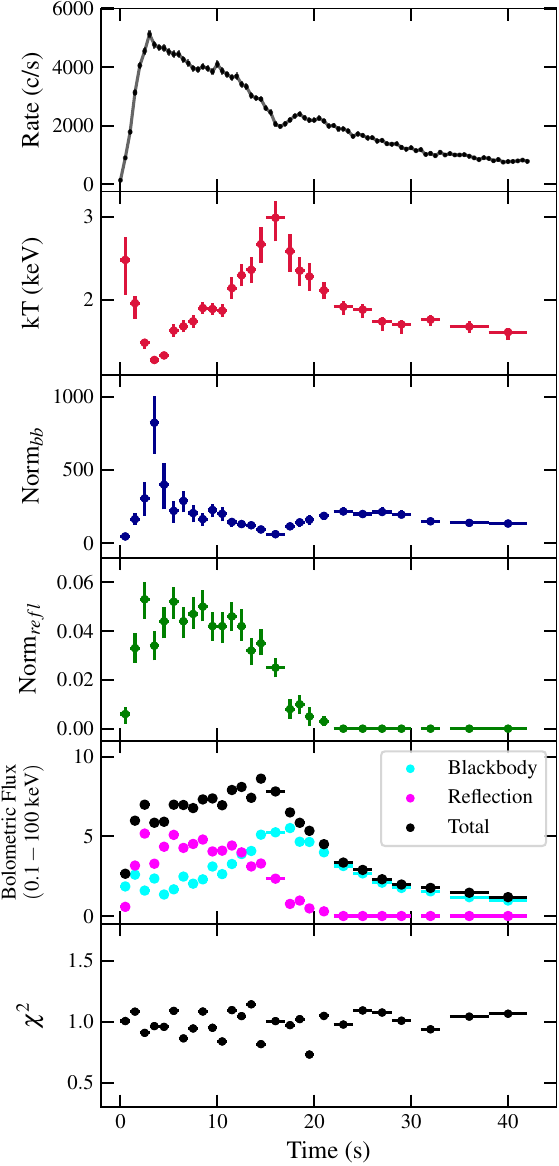}
\caption{Evolution of spectral parameters of \src\ obtained with the disk reflection modeling {\tt relxillNS} approach during the evolution of the thermonuclear burst with \nicer{}. The reported flux values are in the unit of 10$^{-8}$ erg cm$^{-2}$ s$^{-1}$. The X-axis indicates the time from the rise of the burst.}
    \label{fig:time_Resolved_nicer_relxillns}
\end{figure}

\subsection{Time-resolved spectral study of bursts}
\label{time_resolved_spectroscopy}
Time-resolved spectral analysis is performed for each burst to probe the dynamic evolution of different spectral parameters. Initially, pre-burst persistent spectral analysis is performed to account for the contribution from persistent emission. The best-fit pre-burst spectral parameters are used during the time-resolved burst spectral modeling. We extracted pre-burst spectra from the \nicer{} and \xmm{} EPIC-PN observations for an exposure of $\sim$1 ks before the onset of the bursts. The pre-burst spectra can be modeled using an absorbed power law and a blackbody model [XSPEC: {\tt tbabs $\times$ (powerlaw + bbodyrad)}]. The best-fitting pre-burst spectral parameters are presented in Table~\ref{tab:tab_preburst}. The photon index, blackbody temperature, and the hydrogen column density are found to be $\sim$1.9, $\sim$1.2~keV, and $\sim2.4\times 10^{22}$ cm$^{-2}$, respectively, for the pre-burst \nicer{} spectrum. The model provides good fit statistics with a reduced $\chi^{2} $ of $\le1.1$. The total unabsorbed flux in the 0.1-100 keV range is estimated to be $2.4\times 10^{-9}$ erg cm$^{-2}$ s$^{-1}$. The \xmm{} EPIC-PN best-fit pre-burst spectral fitting results for three segments provide the photon index, blackbody temperature and hydrogen column density in the ranges of 1.7-2.1, 1.3-1.6 keV, and (2.0-2.4)$\times10^{22}$ cm$^{-2}$, respectively. The unabsorbed pre-burst flux in the 0.1--100 keV range for these three segments is estimated to be in the range of 6.5-7.0 $\times 10^{-9}$ erg cm$^{-2}$ s$^{-1}$. 

To perform the time-resolved spectral analysis of the thermonuclear burst observed with \nicer, we created a total of 27 spectra (P1-P27). We used a bin size of 1\,s to generate the time-resolved spectra, and the bin size was increased to 2~s and 4~s gradually to optimize the signal-to-noise ratio during the decay phase of the burst. Similarly, to extract time-resolved burst spectra for the bursts observed with \xmm{} EPIC-PN, we used a bin size of 1\,s, and in the decay part, the bin size increases to 2\,s and 4\,s at the tail of the bursts. During the \xmm{} bursts TNB-2 and TNB-3, only the initial $\sim$20 seconds from the onset were detected, capturing the peak and a portion of the cooling tail. The time-resolved burst spectra can be modeled with an absorbed blackbody model. The persistent emission is accounted for by using the best-fit pre-burst value. The combined model in {\tt XSPEC} reads as {\tt tbabs$\times$(bbodyrad1 + powerlaw + bbodyrad2)}, where {\tt bbodyrad1} describes the burst emission, and the model {\tt powerlaw + bbodyrad2} is used for persistent emission. At a fixed value of pre-burst parameters, the \nicer{} burst spectra show residuals in the soft X-ray range (below 2 keV) with reduced $\chi^{2}$ values greater than 1.5. The soft excess emission is possibly due to the enhanced mass accretion rate during the burst, which can be addressed using the variable persistent emission approach (the $f_a$ method, \citet{Worpel2013}). In this method, the persistent emission is allowed to vary during the burst, and a scaling factor is used to accommodate the observed excess emission. The variable persistent emission modeling approach significantly improves the fitting results for \nicer{} burst spectra. We used the F-test to determine whether the $f_a$ component is significant in model fitting. The $f_a$ modeling approach is employed if the chance likelihood of a decrease in $\chi^2$ values is less than 5\%; otherwise, the $f_a$ value is fixed at 1.

The results of the \nicer{} time-resolved burst spectral analysis indicate that during the peak of the burst, the photosphere is expanded above the NS. During the PRE event, the photospheric radius is extended to a maximum value of $23.1_{-3.2}^{+3.8}$ km. While estimating the radius of the NS photosphere, a color correction factor$f_c$ of 1.4 is applied \citep{Suleimanov2011, Va24} and the radius is scaled by the factor $f_{c}^2/(1+z)$, assuming a redshift of 0.3 and the source distance of 5.6 kpc \citep{Ga08}. The corresponding temperature attains a minimum of $1.4\pm0.1$ keV during the PRE phase. However, the uncertainties in measuring the source distance affected the measurements of the actual photospheric radius. A higher source distance of 8.4 kpc was reported for anisotropic burst emission \citep{Va24}. If the source distance is considered as 8.4 kpc, the photospheric radius is estimated to be $34.6_{-4.8}^{+5.8}$ km. 

Figure~\ref{fig:fig_time_Resolved_nicer} shows the evolution of the time-resolved burst spectral parameters using \nicer{} using the variable persistent emission approach (left panel) and constant background modeling approach without any scaling factor (right panel). The top panels show the 0.5-s binned light curve of the burst. The second and third panels show the evolution of blackbody temperature and normalization. For the $f_a$ modeling approach (left panels), at the peak of the burst, the blackbody temperature reached a minimum value of $1.35\pm0.08$ keV, and the corresponding normalization reached a maximum of $746_{-205}^{+245}$, indicating PRE. The variation of the $f_a$ parameters is shown in the fourth panel, and $f_a$ reached a maximum of $26.0\pm3.0$ during the peak of the burst. The evolution of blackbody flux and total unabsorbed flux in the 0.1-100 keV range is shown in the fifth panel. The peak flux reached a value of $(2.6\pm0.1)\times10^{-7}$ erg cm$^{-2}$ s$^{-1}$ at the touchdown phase, which corresponds to the second peak in the blackbody temperature (around 17 s from the burst start time). At the touchdown phase, the photosphere of the NS returned to the surface, and the blackbody temperature reached a maximum value. The peak luminosity reached the Eddington limit. The bottom panel of Figure~\ref{fig:fig_time_Resolved_nicer} shows the corresponding value of reduced $\chi^{2}$ for the best-fit time-resolved spectra of each segment. In the constant background modeling approach (right panels), we found that the dynamic evolution of different spectral parameters, such as blackbody temperature and normalization, exhibits trends similar to those observed in the variable persistent emission approach. However, the blackbody temperature and thus the bolometric flux are comparatively lower than the results from the \( f_a \) modeling approach. We noted excess residuals in constant background modeling, which often lead to a higher reduced chi-square for intense bursts with a larger blackbody expansion radius. Previously, a similar finding was reported in a higher flux value in the constant background approach for different sources \citep{Ja25, Gu22}.

We also performed a detailed time-resolved spectral study of three thermonuclear bursts detected with \xmm{} EPIC-PN using the same approach as \nicer{} spectral modeling. For the modeling of \xmm{} EPIC-PN burst spectra, any additional scaling factor is not required. The burst spectra can be well modeled with an absorbed blackbody model. The bursts do not show any evidence of PRE. The evolution of spectral parameters during these bursts is shown in Figure~\ref{fig:fig_time_Resolved_XMM}. The peak blackbody temperature for TNB-1 is found to be $2.3\pm0.2$ keV, and the corresponding blackbody flux and total unabsorbed flux are $(2.9\pm0.1)\times10^{-8}$ erg cm$^{-2}$ s$^{-1}$ and $(3.5\pm0.1)\times10^{-8}$ erg cm$^{-2}$ s$^{-1}$, respectively, in the 0.1-100 keV range. The spectral parameters during the \xmm{} EPIC-PN TNB-2 and TNB-3 show a similar type of evolution with comparable peak flux. In contrast, in TNB-1, the peak flux is comparatively low, and the burst profile shows a sharp decay compared to the other two bursts. The peak blackbody temperature during TNB-2 is $2.5\pm0.2$ keV, blackbody flux and total flux are estimated to be $(3.5\pm0.2)\times10^{-8}$ erg cm$^{-2}$ s$^{-1}$ and $(4.2\pm0.2)\times10^{-8}$ erg cm$^{-2}$ s$^{-1}$, respectively.
 
\subsubsection{Burst-disk interaction: Disk reflection during the \nicer{} burst}
The time-resolved \nicer{} burst spectra are also modeled following an alternative method, the disk reflection modeling approach. The modeling approach with the $f_a$ method does not provide a detailed physical explanation of the enhanced pre-burst emission. Alternatively, the burst photons may interact with the disk and be reflected. We attempted to use the disk reflection modeling approach to explain the deviation of the burst emission from pure blackbody emission. The disk reflection model {\tt relxillNS} provides a physically motivated and self-consistent explanation. For the spectral modeling of the \nicer{} burst spectra, we utilized the latest relativistic disk reflection model {\tt relxillNS} along with the absorbed blackbody component. In the {\tt relxillNS} model, the photoionized accretion disk is illuminated by a blackbody spectrum from the neutron star \citep{Garcia2022}. The \nicer{} burst spectra can be well modeled using the combined model {\tt tbabs $\times$ (bbodyrad1 + relxillNS + powerlaw + bbodyrad2)}, where {\tt bbodyrad1}, {\tt relxillNS}, and ({\tt powerlaw + bbodyrad2}) describe the contributions of the burst emission, disk reflection contribution, and pre-burst persistent emission, respectively. During fitting of the time-resolved burst spectra using the reflection model, the pre-burst model parameters are fixed at the best-fitted pre-burst values (see Table~\ref{tab:tab_preburst}). The disk reflection model {\tt relxillNS} includes several key parameters such as: disk inclination $i$, inner and outer disk radii $R_{\textrm{in}}$ , $R_{\textrm{out}}$,  dimensionless spin parameter $a$, inner and outer emissivity indices $q1$ and $q2$, iron abundance  $A_{\textrm{Fe}}$, input blackbody temperature $kT_{\textrm{bb}}$, disk ionization parameter $\log\xi$, and disk density $\log{N}$. Since the reflection model consists of many parameters, it is challenging to constrain all model parameters simultaneously in such short time segments. During spectral modeling, we freeze the following parameters at its typical value for similar type of sources: assuming a single emissivity profile $q1=q2=3$, $A_{\textrm{Fe}}$ = 5, $i$ = 40$^{\circ}$, $\log{N}$=19 cm$^{-3}$, $R_{\textrm{in}}$ = $R_{\textrm{ISCO}}$, $R_{\textrm{out}}$ = 400 $R_g$, and $\log\xi$ = 3.2 erg cm s$^{-1}$. The temperature of the input blackbody spectrum in {\tt relxillNS} is tied with the temperature of the {\tt bbodyrad1} component and the reflection fraction for {\tt relxillNS} is set to -1, so that the {\tt bbodyrad1} model describes the direct coronal emission and the {\tt relxillNS} describes the reflection component. The normalization of the reflection model component is allowed to vary during spectral modeling with {\tt relxillNS}. The evolution of the spectral parameters is shown in Figure~\ref{fig:time_Resolved_nicer_relxillns}. The top panel of the figure shows the 0.5-s binned \nicer{} light curve during the burst. The figure illustrates the dynamic evolution of various spectral parameters, including blackbody temperature (second panel), blackbody normalization (third panel), normalization of the {\tt relxillNS} model (fourth panel), and the flux of the blackbody and reflection components (fifth panel). The result also indicates a PRE burst, with the blackbody temperature reaching a minimum of $1.3\pm0.4$ keV and the blackbody normalization attaining a maximum value of $822_{-185}^{+212}$ at the peak of the burst. At the touchdown phase, the blackbody temperature showed a maximum value of $3.0\pm0.2$ keV as the photosphere returns to the NS surface after the photosphere expansion phase. The blackbody flux also reached a maximum of $(5.5\pm0.2)\times10^{-8}$ erg cm$^{-2}$ s$^{-1}$ during the touchdown phase. The flux of the reflection component is  maximum at $(5.2\pm0.2)\times10^{-8}$ erg cm$^{-2}$ s$^{-1}$.

\subsection{Analysis of persistent spectra: Reflection}

 One of the objectives of this work is to investigate the reflection feature in persistent emission. We utilize the recent \nustar{} observation in March 2025 to explore the X-ray reflection feature in the persistent spectrum of \src{}. During the \nustar{} observation, the source did not show any thermonuclear burst. We used the \nustar{} broadband data in the 3--50 keV range for spectral study. The \nustar{} FPMA and FPMB spectra are simultaneously fitted in {\tt XSPEC}, with an additional constant multiplicative factor set at 1 for FPMA and free for FPMB to account for the cross-calibration uncertainties. We used the {\tt TBabs} model \citep{Wilms2000} to consider the absorption along the line of sight caused by the Galactic neutral hydrogen. For the initial input abundance in the {\tt TBabs} model, we employed the {\tt wilm} abundance model. 
 
\begin{table}
	\centering
	\caption{Broadband persistent spectral fitting parameters of \src{} for different disk reflection model combinations. The best-fit is obtained with spectral model \textcolor{black}{M1: \texttt{tbabs $\times$ (bbodyrad + relxill)} and M2: \texttt{tbabs $\times$ (bbodyrad + relxillCp)}}. All errors are calculated using the MCMC in {\tt XSPEC} and are 90\% significant.}
	\label{tab:fitstat2}
	\begin{tabular}{lcc} 
\hline
				
 Parameters	&	 M1& M2	\\	
\hline																			
N$_H$ ($10^{22}$ cm$^{-2}$) 	&	 $2.3^f$ & $2.3^f$  \\					

kT (keV) & $1.80\pm0.04$ & $1.60_{-0.05}^{+0.18}$  \\
Norm	& $1.4\pm0.2$ &	$2.2\pm0.7$	\\
 
Incl (deg) 	&	 $38.9_{-0.4}^{+0.6}$ 	&	$29.5_{-4.2}^{+7.9}$  \\
                   	
$R_{\mathrm{in}}$ ($R_{\mathrm{ISCO}})$	&		 $1.9\pm0.2$ 	&	$1.6_{-0.4}^{+1.0}$  \\
 
$A_{\mathrm{Fe}}$	&	 $2^f$ 	&	$2^f$  \\
q	&		 $3^f$ 	&	 $3^f$  \\
$\log \xi\;(\mathrm{erg~cm~s}^{-1})$ &	 $4.2\pm0.02$ 	&	$4.1_{-0.1}^{+0.2}$ \\
$\log{N}$ (cm$^{-3}$) 		& $-$ 	&	$17.8_{-2.4}^{+0.7}$  \\
$\mathrm{Refl}_{\mathrm{frac}}$	&	 $4.6\pm0.1$ 	&	$0.7_{-0.13}^{+0.6}$  \\
Photon index ($\Gamma$)	& $1.80\pm0.01$ 	& $1.8_{-0.02}^{+0.06}$  \\
Cutoff energy ($E_{\mathrm{cut}}$)	& $137_{-8.9}^{+6.8}$ 	& $-$  \\
kT$_{\rm e}$ (keV) & $-$ & $16.3_{-1.2}^{+2.8}$ \\
Norm ($\times10^{-3})$ 	&$0.70\pm0.03$ 	&	$2.2\pm0.4$ \\
\hline 
Flux$_{0.1-100~\mathrm{keV}}^a$ 	&	 $3.50\pm0.01 $ 	&	$3.50\pm0.01$  \\
                   	
Luminosity$_{0.1-100~\mathrm{keV}}^b$ 	  	&	 $1.34\pm0.05$ 	& $1.34\pm0.05$ \\

\hline                   																
 $\chi^2$/dof  &	1701/1600         &  1699/1599     \\
\hline		
\\
		\multicolumn{3}{l}{$^a$ : Unabsorbed flux in the units of $10^{-9}$ \erg.}\\
    \multicolumn{3}{l}{$^b$ : in the units of $10^{37}$ \lum, for a distance of 5.6 kpc.}\\
      \multicolumn{3}{l}{$^f$ : Frozen parameters.}\\
	\end{tabular}
\end{table} 
\begin{figure}
\centering
\includegraphics[width=0.48\columnwidth,angle=270]{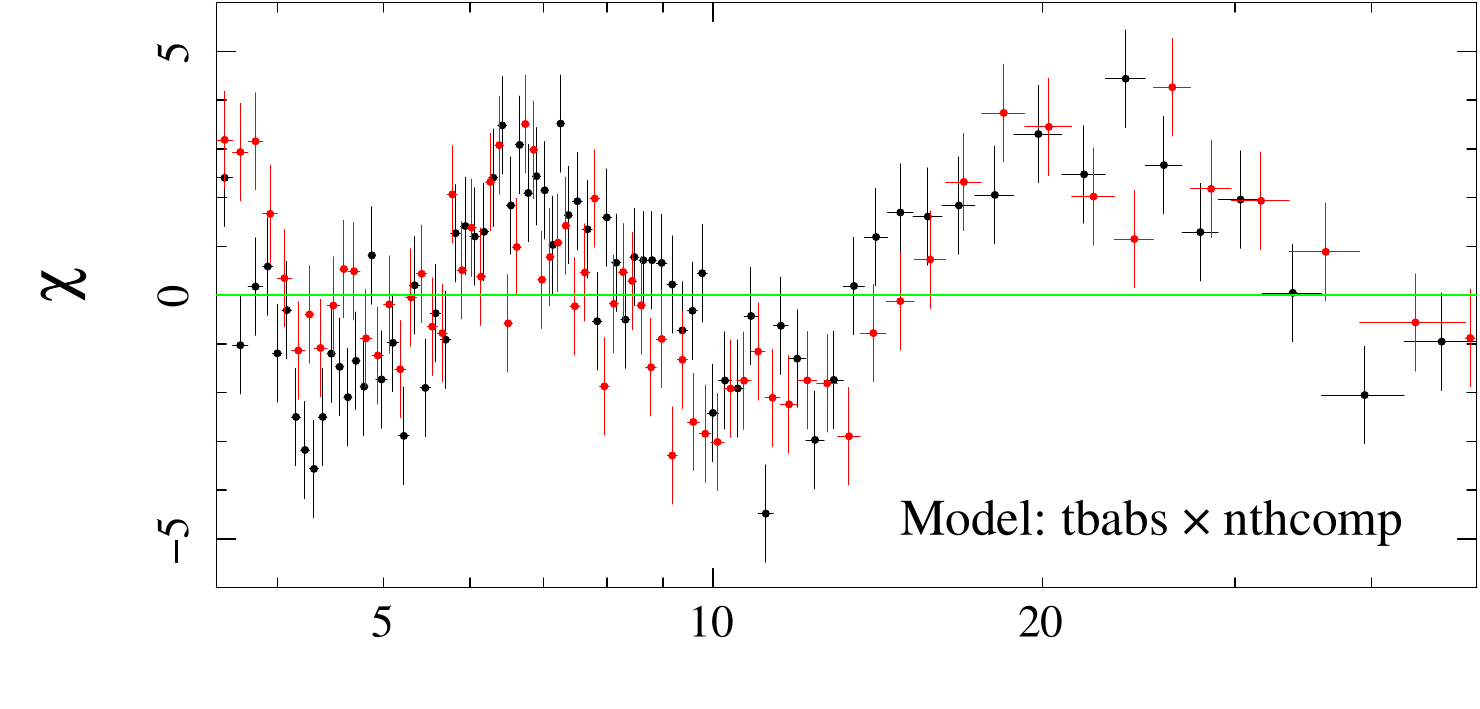}
\includegraphics[width=0.48\columnwidth,angle=270]{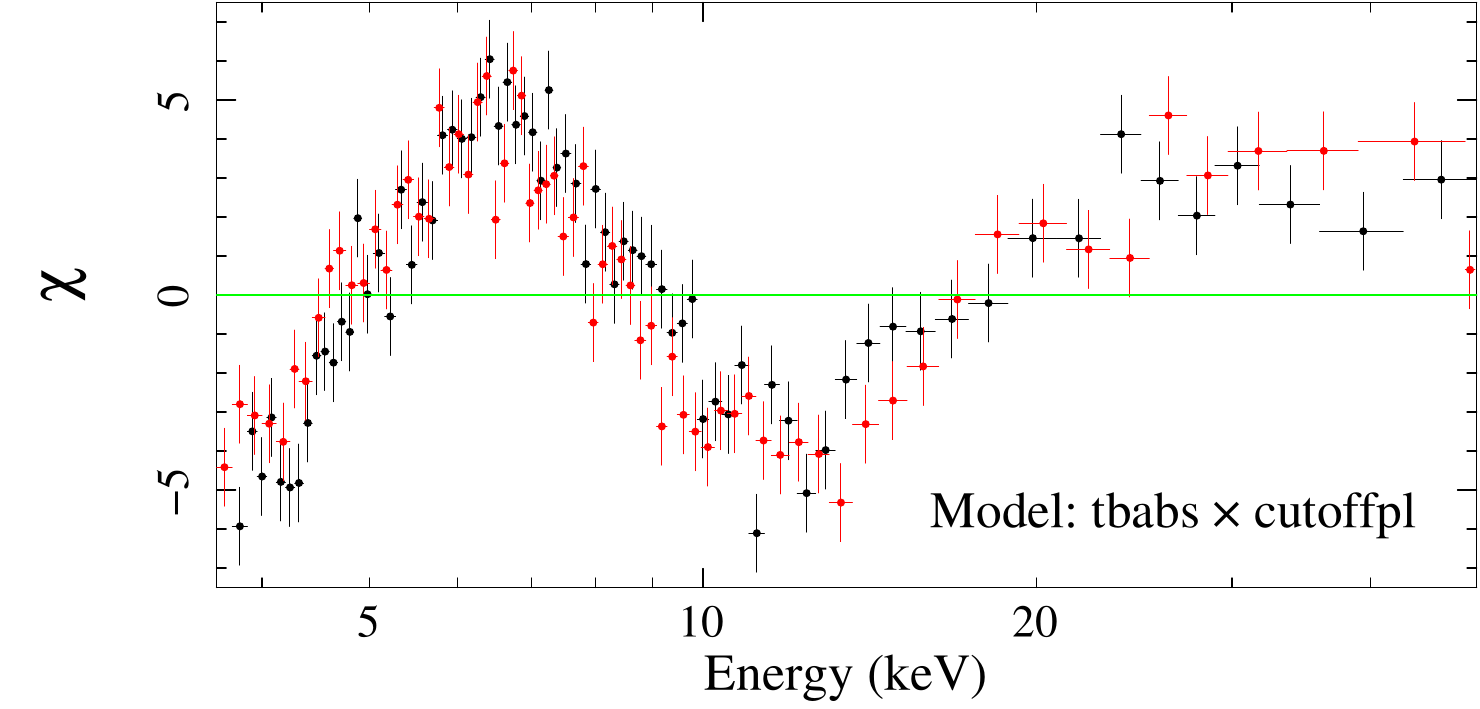}
 \caption{The residuals obtained from the spectral fitting of the persistent emission of \src{} using \nustar{} observation. The top and bottom panels show the residuals obtained by fitting the data with \texttt{tbabs $\times$ nthcomp} model, and  \texttt{tbabs $\times$ cutoffpl} model, respectively. An iron line  $\sim$6.4 keV and a Compton hump at around 20 keV are visible in both the cases. } 
\label{fig:delchi}
\end{figure}

\begin{figure}
\centering
 \includegraphics[width=6cm, angle=270]{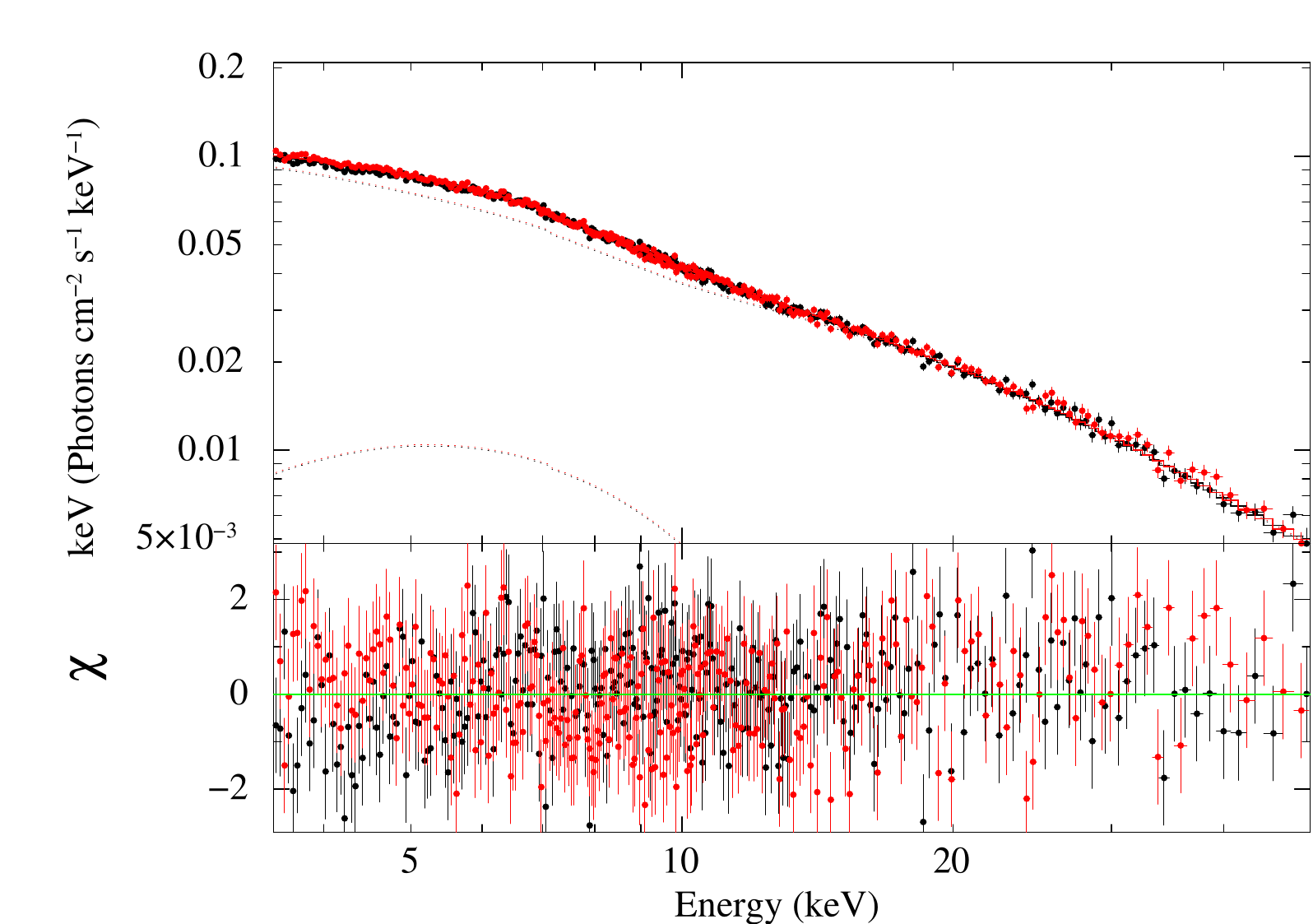}
 \caption{Best-fit \nustar{} spectrum of \src{} with model \texttt{tbabs $\times$ (bbodyrad + relxill)}. The bottom panel represents the residuals obtained from the spectral fit.}
\label{fig:spec_reflection}
\end{figure}

To model the persistent spectrum of \src~ from the \nustar{} observation, we utilize single-component continuum models, such as thermal Comptonization component ({\tt nthcomp}, the multi-temperature disk blackbody component ({\tt diskbb}: \citet{Mi84, Ma86}), single-temperature blackbody component ({\tt bbodyrad}, the cutoff power law model, modified by the {\tt TBabs} model. This single-component modeling approach yields a high value of reduced chi-square ($\chi^2>1.5$). Regardless of the choice of the continuum model, the iron line near 6.4 keV and the Compton hump are visible at around 20 keV. Figure~\ref{fig:delchi} shows the residual plots for two different model combinations {\tt tbabs $\times$ nthcomp} and {\tt tbabs $\times$ cutoffpl} in top and bottom panels, respectively. In both the cases, the residuals show a broad iron line and a Compton hump, indicating X-ray reflection features in the spectrum \citep{Fa89, Fa00, Mi07}. These features typically originate, possibly due to the relativistic reflection from the accretion disk. 

The physically motivated self-consistent relativistic disk reflection model {\tt relxill/rexillCp} is a member of the {\tt relxill} (v2.3: \citet{Ga14, Da14}) model family, which describes relativistic reflection caused by the {\tt nthcomp} or cut-off power-law continuum. The primary source is assumed to have an emissivity given by Index~1 (q1) and Index~2 (q2). The reflection model {\tt relxill} parameters are as follows: the photon index ($\Gamma$) of the illuminating radiation, cutoff energy ($E_{\mathrm{cut}}$), inner disk radius ($R_{\mathrm{in}}$), outer disk radius ($R_{\mathrm{out}}$), disk inclination angle ($i$), dimensionless spin parameter ($a^\ast$), ionization parameter ($\log\xi$) at the surface of the disk, the abundance of Fe ($A_{\mathrm{Fe}}$) relative to its solar value. The model {\tt relxillCp} contains additional variable parameters, such as electron temperature ($kT_e$), and disk density ($\log{N}$). The {\tt relxill} model incorporates a cutoff power-law model as illuminating continuum, whereas the {\tt relxillCp} model includes the thermal Comptonization model {\tt nthcomp} as an illuminating continuum. To model the reflection features in the \nustar{}  spectrum of \src{}, the emissivity index is fixed at q1 = q2 = 3, assuming single emissivity. The outer radius of the accretion disk ($R_{out}$) is fixed at 1000 gravitational radii ($R_g = \frac{GM}{c^2}$). The hydrogen column density is fixed at $2.3 \times 10^{22}~\mathrm{cm^{-2}}$, dimensionless spin parameter $a^\ast$ is set to 0 and the iron abundance is set to twice of the solar abundance ($A_{\mathrm{Fe}} = 2A_{\mathrm{Fe, solar}}$) \citep{Ia16, Lu19}. 

\begin{figure*}
\centering
 \includegraphics[width=0.9\linewidth]{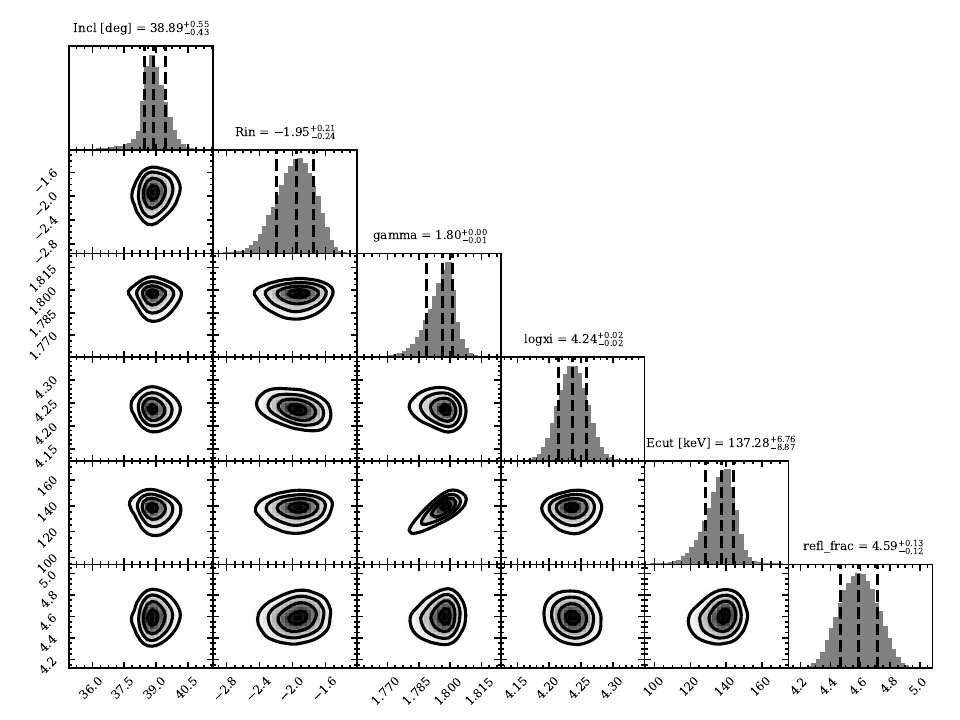}
\includegraphics[width=0.9\linewidth]{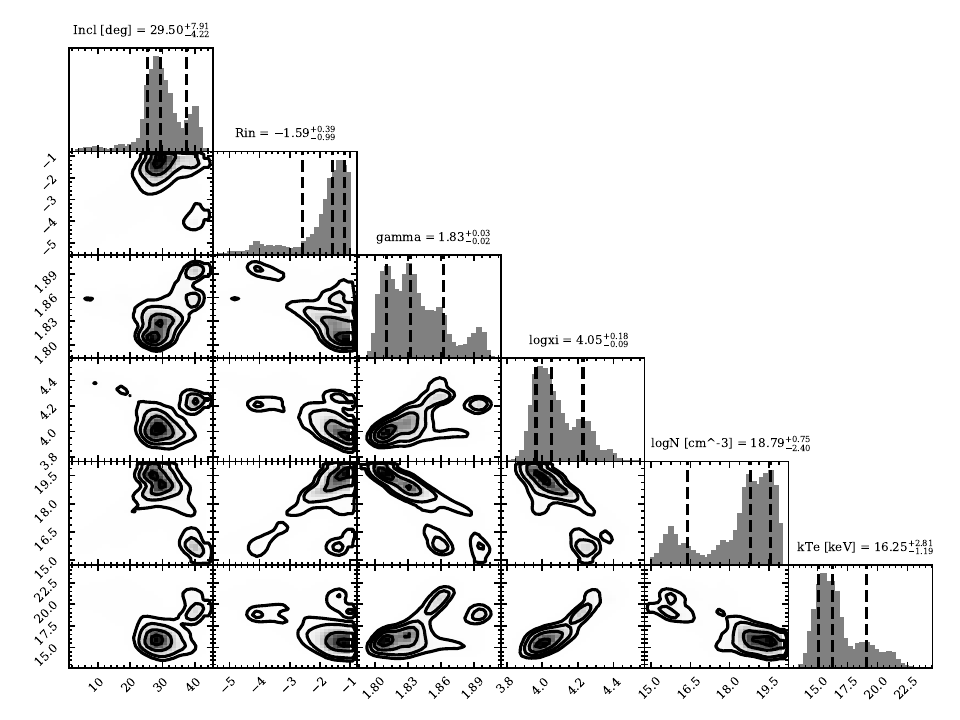}
 \caption{MCMC chain corner plot is shown for the best-fitting spectra with the model-M1 (top panel) and model-M2 (bottom panel).}
\label{fig:corner}
\end{figure*}

We used two different modeling approaches: with model-1 (M1): {\tt TBabs $\times$ (bbodyrad + relxill)} and model-2 (M2): {\tt TBabs $\times$ (bbodyrad + relxillCp)} along with a multiplicative constant. Both models give a significant improvement in the spectral fitting results (reduced $\chi^2\sim$1). Figure~\ref{fig:spec_reflection} presents the best-fit \nustar{} spectra of \src{} with the model M1. The residuals, shown in the bottom panel of Figure~\ref{fig:spec_reflection}, show that the model combination is capable of addressing the reflection features. Table~\ref{tab:fitstat2} provides a summary of the best-fit spectral parameters for models M1 and M2. The best-fit spectral parameters from model M1 provides inner disk radius R$_{in}$, and disk inclination $i$ of $1.9 \pm 0.2~R_{\mathrm{ISCO}}$, and $38.9^{+0.6}_{-0.4}~{}^{\circ}$, respectively. The photon index and cutoff energy are found to be $1.8\pm0.01$ and $137_{-9}^{+7}$ keV, respectively. The {\tt relxillCp} model (model M2) provides the electron temperature of $16.3_{-1.2}^{+2.8}$ keV and the disk density $\log N = 17.8^{+0.7}_{-2.4}~\text{cm}^{-3}$. The disk is found to be strongly ionized in both models with an ionization parameter $\log \xi \sim 4.2~\mathrm{erg~cm~s}^{-1}$. In addition to the reflection model, a blackbody model ({\tt bbodyrad}) is also used to fit the spectrum, which improves the fitting results and returns a blackbody temperature of $\sim1.8$ keV and a relatively lower normalization. The unabsorbed total flux in 0.1-100 keV range is estimated to be $\sim 3.5 \times 10^{-9}~\mathrm{erg~cm^{-2}~s^{-1}}$ and the corresponding luminosity is $\sim 1.34 \times 10^{37}~\mathrm{erg~s^{-1}}$ assuming a source distance of 5.65 kpc. Using the Markov Chain Monte Carlo {\tt MCMC} approach, the reported errors in the spectral parameters are evaluated. Figure~\ref{fig:corner} shows the {\tt MCMC} chain corner plot. Based on the Goodman-Weare algorithm \citep{Good2010}, the {\tt MCMC} simulation is carried out using the reflection models ({\tt relxill/relxillCp}) with a chain length of 500000 and 10 walkers. Since the first 50000 steps were assumed to be in the `burn-in' or `transient' phase, we excluded them from the analysis.

\section{Discussion}
\label{dis}
 In this work, we report the results obtained from a detailed study of thermonuclear X-ray bursts detected from \src{} using \nicer{} and \xmm{} observatories. The profile of the thermonuclear bursts observed with \nicer~ and \xmm~ exhibits a strong energy dependence. Among the four bursts (three with \xmm~ and one with \nicer), the burst observed with \nicer~ is recorded with the highest peak count rate of $\sim$8000 counts s$^{-1}$. A weak secondary peak is also observed in the decay phase of the \nicer~ burst. The weak secondary peak is more prominent in the high energy bands ($\ge3$ keV) than in the low energy band (0.5-3 keV; see Figure~\ref{fig:HR}). The appearance of the secondary peak in the cooling phase is seen in the thermonuclear bursts of several sources during which the photospheric radius expansion event has been reported, such as 4U~1608-52 \citep{Ja19}, GRS 1741.9-2853 \citep{Pike2021}, and SAX~J1808.4-3658 \citep{Bu19}.
 Several possible explanations exist for the observed secondary peak in the burst profile. Burning of the leftover accreted matter on the surface of the neutron star, or reburning of fresh material \citep{Ke17}, may be the cause of the weak secondary peak in the decay phase of the burst. However, the concept of reburning of the fresh accreted matter remains unclear. It is proposed that a hydrodynamical shear instability brought on by convection during the burst might cause new fuel to accumulate on top of the burnt material \citep{Fu88}.
  
A detailed time-resolved spectral study of four bursts is carried out to understand the dynamic evolution of spectral parameters and emission mechanisms. From our analysis, we noticed the occurrence of a PRE event during the intense burst observed with \nicer. The HR profile shows a significant evolution in the spectral state during the PRE burst. The exceptionally high radiation pressure during the intense PRE burst can alter the geometry of the accretion flow, resulting in changes in the spectral shape \citep{Yu24, Ja24}. From our spectral study of the \nicer{} burst, it is found that the burst emission deviated from the pure blackbody emission. This deviation is further investigated using two different approaches, the variable persistent emission modeling approach and the relativistic disk reflection modeling approach, as described in the following subsection. Additionally, we utilize a recent \nustar{} observation to study the reflection feature in the persistent spectrum, which can be used to explore the geometry and characteristics of the accretion disk. 

\subsection{Time-resolved burst spectroscopy and bursting regime}
We perform a detailed study of the \nicer{} thermonuclear burst and three bursts from \xmm{} EPIC-PN. The dynamic evolution of the spectral parameters is investigated during each burst. Among these four bursts, only the \nicer{} burst shows clear evidence of PRE. As the \xmm{} EPIC-PN thermonuclear bursts do not show any evidence of PRE, any additional scaling factor in spectral modeling is not required. The time-resolved \nicer~ burst spectra can be described using the variable persistent emission modeling approach, which provides a more accurate description of the soft excess over the blackbody emission during the burst. The scaling factor ($f_a$) shows a maximum value of $\sim$26 during the peak of the burst. During the \nicer~ burst, the peak flux reached the Eddington limit, and the photosphere expanded to a maximum of $23.1_{-3.2}^{+3.8}$ km. The corresponding blackbody temperature reached a minimum value of $1.4\pm0.1$ keV at the peak. After the expansion phase, the NS photosphere returned to the NS surface, and the temperature again increased to a maximum of $5.6_{-3.2}^{+1.4}$ keV, corresponding to the touchdown phase. The burst fluence for the \nicer{} burst is estimated to be $(4.2\pm0.1) \times 10^{-6}$ erg cm$^{-2}$ using the total blackbody flux integrated over each time segment. The corresponding net energy release during the burst is estimated to be $(1.5\pm0.1)\times 10^{40}$ erg. The burst spectrum deviates from a pure blackbody emission or the Planck function during intense X-ray bursts, as observed in \src~ during the PRE burst (present work). The deviation of the burst emission can be explained using different approaches, including the NS atmospheric effect \citep{Ozel2013}, which predicts a comparatively broad X-ray spectrum instead of a pure blackbody \citep{Suleimanov2011}. The enhanced mass accretion rate due to the Poynting-Robertson drag \citep{Walker1992, Worpel2013} can also be expected during the burst, and the extreme radiation drag may drain out the inner accretion disk onto the NS \citep{De18}. 

During an intense X-ray burst, the reflection of burst photons from the accretion disk is also anticipated to explain the soft spectral excess emission \citep{Ba04, Ma25, SRGA25}. This approach was adopted to describe the burst spectra for different sources such as Aql~X-1 \citep{Ma25}, 4U~1820--30 \citep{Ke18b, Ja24}, 4U~1730--22 \citep{Lu23}, 4U~1636--536 \citep{Zh22}, SAX~1808.4--3658 \citep{Bu21}, etc. We made an effort to examine the impact that disk reflection had during the bursts in this context. The disk reflection model {\tt relxillNS} can be used to describe the soft excess emission during the burst. The PRE event in \src~ is evident irrespective of the modeling approach. In the disk reflection modeling approach, the NS photosphere is extended to a maximum of $24.2_{-2.7}^{+3.1}$ km, and the corresponding temperature reaches a minimum of $1.3\pm0.4$ keV. After the PRE phase, the blackbody temperature started to increase and reached a maximum of $3.0\pm0.2$ keV at the touchdown phase. The $f_a$ modeling approach provides a comparatively higher blackbody temperature and flux than the relativistic disk reflection model. The reflection model yielded a similar blackbody radius and a lower blackbody temperature, resulting in a lower bolometric flux compared to the variable persistent emission method. Similar findings are also observed for other sources (e.g., 4U~1730-22, \citet{Lu23}). The accretion disk may block and/or enhance the observed burst flux when the burst photons are reflected. The disk geometry and inclination angle with respect to the line of sight influence the anisotropy of the burst emission \citep{He16}.

 We also investigated the ignition regime and the burning mechanism during the thermonuclear burst. We estimated the mass accretion rate, ignition column depth, and burst fluence. The thermonuclear ignition regime can be investigated based on the local mass accretion rate onto the NS. The mathematical expression for the local accretion rate onto the NS is given as \citet{Ga08}:

\begin{align}
\dot{m} =\; & \frac{L (1+z)}{4\pi R^2 (GM/R)} \nonumber \\
=\; & 6.7 \times 10^{3} 
\left( \frac{F_{b}}{10^{-9}~\mathrm{erg~cm^{-2}~s^{-1}}} \right) 
\left( \frac{d}{10~\mathrm{kpc}} \right)^2 
\left( \frac{1+z}{1.31} \right) \nonumber \\
& \times \left( \frac{R}{10~\mathrm{km}} \right)^{-1} 
\left( \frac{M}{1.4~M_\odot} \right)^{-1} 
~\mathrm{g~cm^{-2}~s^{-1}}
\end{align}

 where $M$, $R$, $d$, and $F_b$ are the mass, radius, distance, and the persistent bolometric flux of the neutron star, respectively. We assume a typical NS of mass of 1.4$~M_\odot$, radius 10 km, gravitational redshift ($z$) of 0.3, and distance to the NS of 5.6 kpc. Following these assumptions, the estimated mass accretion rate is $0.5~\times~10^{4}$ g~cm$^{-2}$~s$^{-1}$ corresponding to $F_b$ = $2.4~\times~10^{-9}$ erg~cm$^{-2}$~s$^{-1}$. For a typical NS, the local Eddington mass accretion rate is $\dot{m}_{\mathrm{Edd}} = 8.8~\times~10^4~\mathrm{g~cm^{-2}~s^{-1}}$. The local accretion rate is related to the Eddington rate as $\dot{m} = 0.06~\dot{m}_{\mathrm{Edd}}$, which is higher ($\ge0.01$) than the required rate for stable burning of hydrogen into helium. 
 The estimated ratio ($\frac{\dot{m}}{\dot{m}_{\mathrm{Edd}}}$) of the mass accretion rate for \src{} with \nicer{} lies between 0.01-0.1 range (see Table~1, \citet{Ga08}). It suggests that in such high pressure and temperature, the helium burning is highly unstable \citep{Ga08}. 
 Similarly, for \xmm{} EPIC-PN thermonuclear bursts, we also estimated the mass accretion rate using the same procedure. The estimated mass accretion rates during the pre-burst persistent emission are found to be between $(1.3-1.6) \times 10^ {4}$ g cm$^{-2}$ s$^{-1}$. The ratio of mass accretion rate to the Eddington mass accretion rate ($\frac{\dot{m}}{\dot{m}_{\mathrm{Edd}}}$) is between 0.15-0.18, which is higher than the \nicer{} measurements.

The column depth ($y_{\textrm{ign}}$) is also calculated at which the burst was ignited by using the bolometric burst fluence following \citet{Ga08}:

   \begin{align}
y_{\textrm{ign}} =\; & \frac{L_b (1+z)}{4\pi R^2 Q_{\textrm{nuc}}} 
\end{align}

 where, $L_{\textrm{b}} = 4 \pi d^{2} E_{\textrm{b}}$ is the bolometric burst luminosity integrated over the burst, $Q_{\textrm{nuc}}$ is the generated nuclear energy, which depends on the fuel composition. For pure helium, the energy released $Q_{\textrm{nuc}}$ =  1.31 MeV nucleon$^{-1}$ (hydrogen mass fraction X=0, \citet{Go19}). Assuming a typical NS of radius 10 km and source distance as 5.6 kpc, the estimated ignition column depth is $y_{\textrm{ign}} = 1.3\times10^{9}$ g cm$^{-2}$ corresponding to the burst fluence of $4.2\times10^{-6}$ erg cm$^{-2}$ in \nicer{} burst. Typically, the unstable burning of helium via the triple $\alpha$ process is comparatively faster than the unstable burning of mixed hydrogen and helium. The burst light curve shows a sharp rise of $\sim3$~s and the exponential decay of $\sim20$ s (by using $\tau=E_\textrm{b}/E{\textrm{peak}}$). The bursts may also originate from a mixed hydrogen/helium fuel composition, with a relatively low hydrogen fraction. The relatively low accretion rate and deep ignition column suggest that the fuel composition is possibly predominantly helium. However, a longer burst duration of more than 30 seconds indicates the possible presence of hydrogen, which is also consistent with mixed hydrogen and helium burning. Depending on the time to accumulate the critical column of the fuel, the burst may be a `mixed H/He' or `pure He' burst \citep{Cu04}.

\subsection{X-ray reflection in persistent emission}
The persistent spectrum of \src~ with \nustar{} exhibits reflection features, including a broad iron line and a Compton hump. These features are investigated using the relativistic disk reflection model {\tt relxill/relxillCp}. From the spectral fitting with {\tt relxill} model, the inner disk radius $R_{\textrm{in}}$ and the disk inclination angle are estimated to be $\sim 2R_{\textrm{ISCO}}$, disk inclination angle $\sim39^{\circ}$. The disk is found to be highly ionized and the disk density is $\log N \sim 18~\text{cm}^{-3}$. Our estimation of the disk inclination angle closely agrees with the earlier findings ($i\sim38^{\circ}$: \citet{Ma19, Ia16}). \citet{Lu19} reported a high iron abundance and high disk ionization using \relxill{} model. However, another study by \citet{Ba24} provides a detailed investigation of the spectral modeling of the 2017 \nustar{} data of \src~ with different reflection models and assumptions. \citet{Ba24} added a blackbody component with the \relxill{} model, which improves the fitting results significantly. \citet{Ba24} concluded from their different approaches of spectral modeling that both the disk ionization and iron abundance are high ($\log \xi > 3.5, A_{\textrm{Fe}} > 1.9 A_{\textrm{Fe, solar}}$) with the \relxill{} modeling approach. The inner disk truncated at $R_{\textrm{in}} \ge 2R_{\textrm{ISCO}}$ in both {\tt relline} and {\tt relxill} models and the disk inclination considered to be 38$^{\circ}$ following \citet{Ma19}. 

The spectral fitting of the \nustar{} data provides the inner disk radius of $\sim2$R$_{\textrm{ISCO}}$. For a rotating neutron star with dimensionless spin parameters $a^\ast$, $\rm R_{ISCO}$ is defined as \citet{Va04},
\begin{align}
R_{\rm ISCO} &= \frac{6GM}{c^2} \left(1 - 0.54 a^\ast \right) 
= 6 \, \frac{GM}{c^2} = 6\, R_g
\end{align}

where $G$, $M$, and $c$ are the universal gravitational constant, mass of the neutron star, and the velocity of light, respectively. The inner disk radius is found to be $R_{\textrm{in}}\sim 2.0~R_{\textrm{ISCO}} \sim 12~R_\textrm{g}\sim25\textrm{~km}$. The inner disk radius in the accreting millisecond X-ray pulsars (AMXPs) is usually less than 15 $R_\textrm{g}$ \citep{Pa09}, although a greater value ($\sim40 R_\textrm{g}$) has been observed in some sources \citep{Pa10, Pa13a}.
 
In the case of NS LMXBs, the accretion disk may be truncated at the magnetosphere boundary due to the magnetic field of the NS \citep{Ca09}. It is possible to estimate the magnetic field strength assuming that the disk is truncated at the magnetosphere radius. The inner disk radius and bolometric flux value from the reflection spectral modeling are utilized for this calculation. The magnetic dipole moment ($\mu$) and the magnetic field can be estimated using the following relation from \citet{Ib09, Ca09}, 

\begin{align}
\mu =\; & 3.5 \times 10^{23} \, x^{7/4} \, k_A^{-7/4} 
\left( \frac{M}{1.4\, M_\odot} \right)^2 \nonumber \\
& \times \left( \frac{f_{\rm ang}}{\eta} \frac{F_b}{10^{-9}\, \mathrm{erg\, cm^{-2}\, s^{-1}}} \right)^{1/2}
\left( \frac{D}{3.5\, \mathrm{kpc}} \right)
\end{align}

where, $\eta$ is the accretion efficiency, $\rm f_{ang}$ is anisotropy correction factor, and $k_A$ is the geometric coefficient. The scaling factor $x$ can be estimated to be $\sim$12 from $\text R_{\rm in} = \frac{xGM}{c^2}$. The 0.1--100 keV flux used as bolometric flux, which is $\text F_{b} \sim 3.5 \times 10^{-9}$ erg cm$^{-2}$ s$^{-1}$. For further calculation of the magnetic field, we assumed $f_{\rm ang} = 1$, $\eta = 0.1$, and $k_A = 1$ \citep{Ca09}. The magnetic dipole moment of the NS can be estimated to be $\mu = 2.56 \times 10^{26}$ G cm$^3$. The corresponding magnetic field strength at the poles of NS of radius 10 km is $B~\sim5.1 \times 10^8$ G.  The estimated magnetic field of \src{} is consistent with the typical value of the magnetic field of other NS LMXBs \citep{Mu15, Lu16}.

\section{Summary and conclusions}
\label{con}
We report a PRE thermonuclear X-ray burst from \src{} with \nicer{}. The photospheric radius is expanded to a maximum of $23.1_{-3.2}^{+3.8}$ km, and the corresponding temperature drops to a value of $\sim$1.4 keV during the PRE. In an intense PRE burst, the accretion flow dynamics may be affected, which can be linked to the changes in the spectral shape. The time-resolved spectra of the \nicer{} bursts can be well modeled with a variable persistent emission approach, suggesting that an increased accretion rate may be influenced by Poynting-Robertson drag. Alternatively, the burst spectra can be well modeled using the disk reflection model {\tt relxillNS}, indicating the burst photons may interact with the accretion disk and be reflected. The pre-burst mass accretion rate is $\sim$6\% of the Eddington mass accretion rate during \nicer{} observation. The PRE burst observed with \nicer~ may be a He-powered or mixed H/He burst. Additionally, we conduct a detailed time-resolved spectral study of three thermonuclear bursts detected by \xmm{} EPIC-PN. The EPIC-PN burst spectra did not show any evidence of PRE, and the time-resolved spectra can be well modeled using an absorbed blackbody model without any scaling factor. We also presented the findings on the X-ray reflection feature in persistent emission from \src{} using a recent \nustar{} observation in March 2025, highlighting an iron line at 6.4 keV and a Compton hump at $\sim$20 keV. The relativistic disk reflection model is utilized to investigate the reflection feature. The best-fit results from the {\tt relxill} model provide an inner disk radius of  $\sim$12$R_g$ and disk inclination angle of $\sim39^{\circ}$. The magnetic field strength is estimated to be $5.1\times10^{8}$ G, assuming the accretion disk is truncated at the magnetosphere boundary.

\facilities{ADS, HEASARC, \nicer{}, \xmm{}, \nustar{}}

\software{HEASoft V6.31.1\footnote{http://heasarc.gsfc.nasa.gov/ftools} \citep{heasoft}, XSPEC V12.13.0 \citep{Ar96}}, NumPy and SciPy \citep{virtanen20}, Matplotlib \citep{hunter07}, IPython \citep{perez07}.

\section*{Acknowledgements}
The research work at the Physical Research Laboratory, Ahmedabad, is funded by the Department of Space, Government of India. This research has made use of data obtained with \nustar{}, a project led by Caltech, funded by NASA, and managed by NASA/JPL, and has utilized the {\tt NUSTARDAS} software package, jointly developed by the ASDC (Italy) and Caltech (USA). We acknowledge the use of public data from the \xmm{} and \nicer{} data archives. We thank the \nustar{} SOC team for making this ToO observation possible. 
\section*{Data Availability}
The data used for this article are publicly available in the High Energy Astrophysics Science Archive Research Centre (HEASARC)\footnote{https://heasarc.gsfc.nasa.gov/db-perl/W3Browse/w3browse.pl}.

\bibliography{4U1702_429_ApJ}{}
\bibliographystyle{aasjournal}
\end{document}